\documentclass[a4paper, 10pt, conference]{ieeeconf}       

\IEEEoverridecommandlockouts                              

\usepackage{cite}
\usepackage{amsmath,amssymb,amsfonts}
\usepackage{makecell,mathtools}
\usepackage{algorithmic}
\usepackage{graphicx}
\usepackage{textcomp}
\usepackage{xcolor}

\title{\LARGE \bf
Subspace Based Identification of Errors-in-Variables Linear Descriptor Systems
}

\author{Deepanjhan Das$^{1}$ and Shankar Narasimhan$^{2}$
\thanks{$^{1}$Deepanjhan Das is with the Department of Chemical Engineering, Indian Institute of Technology Madras, Chennai 600036, India {\tt\small deepanjhan.iitm22@gmail.com}}%
\thanks{$^{2}$Shankar Narasimhan is a Professor Emeritus in the Department of Chemical Engineering, Indian Institute of Technology Madras, Chennai 600036, India {\tt\small naras@iitm.ac.in}}%
}

\begin{document}
\maketitle
\thispagestyle{empty}
\pagestyle{empty}

\begin{abstract}
The identification of linear descriptor systems (DAEs) from noise-corrupted data makes two critical assumptions: requirement of an \textit{a priori} classification of variables into inputs and outputs, and a pre-specified structural assumption with respect to the index of the system. This paper proposes a data-driven methodology for identifying index-0 and index-1 DAEs within an errors-in-variables framework. We extend a subspace-based iterative PCA (SMI-IPCA) approach to the behavioral setting, treating all measured variables as a unified augmented vector to avoid classification bias. This method enables systematic estimation of the noise variances, the number of algebraic and differential output variables, while simultaneously identifying the algebraic constraints and kernel representation of the dynamic system corresponding to its minimal realization order without prior structural knowledge. Simulation studies on index-0 and index-1 systems demonstrate the effectiveness of the proposed approach and its practical applicability.
\end{abstract}

\section{INTRODUCTION} \label{sec:intro}
Linear descriptor systems, governed by coupled set of differential and algebraic equations, arise naturally in a wide range of engineering applications, including electrical circuit networks, mechanical multibody systems, chemical process plants \cite{Kunkel:2024}. Unlike standard state-space models, the presence of algebraic constraints in descriptor systems encode physical laws and conservation principles directly within the model structure \cite{Kunkel:2024,Dai:1989}. Consequently, acquiring accurate data-driven models of such systems from measured data is of significant practical importance for downstream tasks such as state estimation, fault detection, and model predictive control.

The classical subspace-based model identification (SMI) methods, such as MOESP~\cite{Verhaegen:1992} and N4SID~\cite{Van:1994} provide an attractive framework for identifying linear multi-input multi-output (MIMO) state-space models directly from data, circumventing the nonlinear optimization required by prediction error methods~\cite{Ljung:2002}. These methods leverage the geometry of the observability subspace via singular value decomposition (SVD) to jointly determine the model order and system matrices. An early extension of subspace identification to descriptor systems was proposed by~\cite{Moonen:1992}, projecting future outputs onto input-output subspaces to extract differential and algebraic components through least-squares estimation. However, a critical limitation shared by all these methods is the assumption that input measurements are noise-free. 

When both inputs and outputs are noisy, the problem is cast as errors-in-variables (EIV) identification. Due to noise in the inputs, standard SMI methods yield biased estimates in the EIV setting. Several subspace-based methods have been proposed to address this, primarily through instrumental variable (IV) approaches~\cite{Chou:1997,Wang:2002}, which neutralize noise effects by using past lagged data as instruments. More recently, SMI-IPCA~\cite{Ramnath:2023} was proposed, which combines SMI with iterative principal component analysis (IPCA) to simultaneously estimate the system order, measurement noise variances, and model matrices.

Despite these advances in EIV identification, two deeply embedded assumptions persist in the existing literature. The first is the \textit{a priori} classification of measured variables into input and output variables, a prejudice that, as noted by Kalman~\cite{Kalman:1982} and reinforced by others~\cite{Los:1989}, may not reflect the underlying physics and imposes an artificial asymmetry inconsistent with the behavioral framework~\cite{Willems:2007}. The second is the \textit{a priori} structural pre-specification of the DAE index, requiring the practitioner to declare whether the system is index-0 (standard ODE) or index-1 (with algebraic constraints) before identification begins. Such structural pre-knowledge is often unavailable in practice and may itself be a source of modeling error.

In this work, we propose a subspace-based EIV identification framework for linear descriptor systems that systematically eliminates both prejudices. The proposed method (i) unifies all measured signals symmetrically, deferring the input-output partition to be inferred from data; (ii) automatically determines whether the system is index-0 or index-1 by detecting the presence of algebraic constraints; (iii) simultaneously estimates measurement noise variances, minimal system order, the number of algebraic and differential/dynamic constraints along with their corresponding implicit constraint matrices (kernels), all from a single identification procedure. Simulation results on an index-0 and an index-1 benchmark system validate the effectiveness of the proposed framework.

 \section{SYSTEM DESCRIPTION AND PROBLEM FORMULATION} \label{sec:sys-desc}
In this section, we proceed to formally describe the class of linear descriptor systems under consideration and the fundamental assumptions made. The relevant notation used throughout this work is introduced, followed by the precise formulation of the identification problem.

\subsection{State-space form of index-0 and index-1 linear DAEs} \label{subsec:2.ss}
Descriptor system, often referred to as a DAE system, generalizes the standard system of ODEs by coupling differential and algebraic equations \cite{Dai:1989}. In this work, we consider a discrete-time, linear time-invariant (LTI) descriptor system, the state-space model of which, in the absence of process noise, can in general be written as:
\begin{subequations}
    \label{eq:2.1}
    \begin{align}
        \mathbf{Ex}(t+1) &= \mathbf{Ax}(t) + \mathbf{Bu}^*(t) \label{eq:2.1a} \\
        \mathbf{y}^*(t) &= \mathbf{Cx}(t) + \mathbf{Du}^*(t) \label{eq:2.1b}
    \end{align}
\end{subequations}
where $\mathbf{x}(t) \in \mathbb{R}^n$ is the state vector, and $\mathbf{u}^*(t) \in \mathbb{R}^\ell$, $\mathbf{y}^*(t) \in \mathbb{R}^m$ represent the true values of the input and output variables, respectively. Throughout this work, we assume the inputs to be random binary signals (RBS). The analysis of descriptor systems requires careful consideration of their structural properties. For such discrete systems, the structural coupling between differential and algebraic constraints is characterized by the properties of the matrix pencil $(s\mathbf{E}-\mathbf{A})$ and its nilpotency index \cite{Luenberger:1978, Dai:1989}, which dictates the minimum number of times part or all of the system equations must be shifted forward in time to obtain an explicit difference equation for $\mathbf{x}(t+1)$ as a function of $\mathbf{x}(t)$ and $\mathbf{u}^*(t)$.

An index-$0$ DAE is equivalent to a standard ODE, which is a special case of general DAE with a non-singular, often identity, leading matrix $\mathbf{E}$, i.e., $\mathbf{E} = \mathbf{I}_n$ in \eqref{eq:2.1a}. For index-$1$ DAE systems, $\mathbf{E}$ is singular with rank$(\mathbf{E}) = n_d < n$, and $\exists$ a non-singular matrix $\mathbf{Q}$ that transforms the states as $\mathbf{x}(t) = \mathbf{Q\tilde{x}}(t)$, where $\mathbf{\tilde{x}}(t) = \begin{bmatrix} \mathbf{x}_d^\intercal & \mathbf{x}_a^\intercal \end{bmatrix}^\intercal$. Consequently, \eqref{eq:2.1a} can be written in the following structured descriptor form:
\begin{subequations}
    \label{eq:2.2}
    \begin{align}
        \mathbf{PEQ\tilde{x}}(t+1)\! &= \mathbf{PAQ\tilde{x}}(t) + \mathbf{PBu}^*(t) \label{eq:2.2a} \\
        \implies\! \begin{bmatrix} \mathbf{x}_d(t\!+\!1) \\ \mathbf{0} \end{bmatrix}\! &=\! \begin{bmatrix} \mathbf{A}_{dd} \!&\! \mathbf{A}_{da} \\ \mathbf{A}_{ad} \!&\! \mathbf{A}_{aa} \end{bmatrix}\!\! \begin{bmatrix} \mathbf{x}_d(t) \\ \mathbf{x}_a(t) \end{bmatrix}\! +\! \begin{bmatrix} \mathbf{B}_d \\ \mathbf{B}_a \end{bmatrix}\! \mathbf{u}^*(t)  \label{eq:2.2b}
    \end{align}
\end{subequations}
where $\mathbf{P}$ is a non-singular transformation matrix with $\mathbf{PEQ} = \mathrm{diag}([\mathbf{I}_{n_d},\ \mathbf{0}])$, and \eqref{eq:2.1b} is rewritten as:
\begin{subequations}
    \label{eq:2.3}
    \begin{align}
        \mathbf{y}^*(t) &= \mathbf{CQ\tilde{x}}(t) + \mathbf{Du}^*(t) \label{eq:2.3a} \\
        \implies \mathbf{y}^*(t) &= \begin{bmatrix} \mathbf{C}_{d} \!&\! \mathbf{C}_{a} \end{bmatrix} \begin{bmatrix} \mathbf{x}_d(t) \\ \mathbf{x}_a(t) \end{bmatrix} + \mathbf{Du}^*(t)
    \end{align}
\end{subequations}
The differential and algebraic components are denoted by subscripts $d$ and $a$, respectively: $\mathbf{x}_d(t) \in \mathbb{R}^{n_d}$ and $\mathbf{x}_a(t) \in \mathbb{R}^{n_a}$ are system states (with $n_d + n_a = n$). For an index-1 DAE system, the submatrix $\mathbf{A}_{aa}$ in \eqref{eq:2.2b} is invertible.  Therefore, we can represent \eqref{eq:2.2} and \eqref{eq:2.3} in terms of only the differential states $\mathbf{x}_d$ by replacing the algebraic states $\mathbf{x}_a$ using \eqref{eq:2.2b} to obtain reduced standard state-space system representation:
\begin{subequations}
    \label{eq:2.4}
    \begin{align}
        \mathbf{x}_d(t+1) &= \mathbf{A}_r\mathbf{x}_d(t) + \mathbf{B}_r\mathbf{u}^*(t) \label{eq:2.4a} \\
        \mathbf{y}^*(t) &= \mathbf{C}_r\mathbf{x}_d(t) + \mathbf{D}_r \mathbf{u}^*(t) \label{eq:2.4b}
    \end{align}
\end{subequations}
where the reduced system matrices are defined as:
\begin{align*}
        &\mathbf{A}_r = \mathbf{A}_{dd} - \mathbf{A}_{da}\mathbf{A}_{aa}^{-1}\mathbf{A}_{ad}; \hspace{0.5em} \mathbf{B}_r = \mathbf{B}_d - \mathbf{A}_{da}\mathbf{A}_{aa}^{-1}\mathbf{B}_{a}; \nonumber \\
        &\mathbf{C}_{r} = \mathbf{C}_{d} - \mathbf{C}_{a}\mathbf{A}_{aa}^{-1}\mathbf{A}_{ad}; \hspace{0.5em} \mathbf{D}_{r} = \mathbf{D} - \mathbf{C}_{a}\mathbf{A}_{aa}^{-1}\mathbf{B}_{a}; \nonumber 
\end{align*}

However, the characterization in \eqref{eq:2.4} treat input and output variables separately, relying on an \textit{a priori} $\mathbf{y}/\mathbf{u}$ classification, which is a modeling prejudice \cite{Kalman:1982, Los:1989} inconsistent with the behavioral framework \cite{Willems:2007}. We therefore adopt a symmetric formulation by unifying all measured signals into an augmented sequence $\mathbf{z}^{*}(t) = \mathbf{\Pi}[\mathbf{y}^*(t)^\intercal,\ \mathbf{u}^*(t)^\intercal]^\intercal$, where $\mathbf{\Pi} \in \mathcal{P}_{n_z}$ is an $n_z\times n_z$ permutation matrix with unspecified variable ordering, and $\mathbf{z}^* \in \mathbb{R}^{n_z}$, i.e., $n_z=m+\ell$. Therefore, in terms of the $\mathbf{z}^*(t)$ vector, \eqref{eq:2.4} can be written as:
\begin{subequations}
    \label{eq:2.5}
    \begin{align}
        \mathbf{x}_d(t+1) &= \mathbf{A}_r^{\mathbf{\Pi}}\mathbf{x}_d(t) + \mathbf{B}_r^{\mathbf{\Pi}} \mathbf{z}^*(t) \label{eq:2.5a} \\
        \mathbf{D}_r^{\mathbf{\Pi}} \mathbf{z}^*(t) &= \mathbf{C}_r^{\mathbf{\Pi}} \mathbf{x}_d(t) \label{eq:2.5b} 
    \end{align}
\end{subequations}
where $\mathbf{A}_r^{\mathbf{\Pi}} = \mathbf{A}_r$, $\mathbf{B}_r^{\mathbf{\Pi}} = \begin{bmatrix}\mathbf{0}_{n_d\times m} & \mathbf{B}_r\end{bmatrix}\mathbf{\Pi}^{-1}$, $\mathbf{C}_r^{\mathbf{\Pi}} = \mathbf{C}_r$, and $\mathbf{D}_r^{\Pi} = \begin{bmatrix} \mathbf{I}_m & -\mathbf{D}_r \end{bmatrix}\mathbf{\Pi}^{-1}$. Based on this, we introduce the dynamic constraint matrix of minimal order $n_d$ as $\boldsymbol{\mathcal{A}}_{d,n_d} = \mathcal{B}(\mathbf{A}_r^{\mathbf{\Pi}},\mathbf{B}_r^{\mathbf{\Pi}},\mathbf{C}_r^{\mathbf{\Pi}},\mathbf{D}_r^{\mathbf{\Pi}})$, a function of the reduced system matrices, that provides the complete description of the dynamic behavior, also known as the kernel representation \cite{Willems:1986,Willems:2007}, while also encoding the algebraic constraints. We formally define the structure of $\boldsymbol{\mathcal{A}}_{d,n_d}$ in Section~\ref{subsec:4.noise}. Let the rank of $\mathbf{C}_r$ in \eqref{eq:2.4b} be $m_d$ and $m_a = m - m_d$. Pre-multiplying both sides of \eqref{eq:2.5b} with the orthogonal complement of $\mathbf{C}_r^{\mathbf{\Pi}}$ we obtain:
\begin{equation}
    \label{eq:2.6}
    \mathbf{W} \mathbf{D}_r^{\mathbf{\Pi}} \mathbf{z}^*(t) = \boldsymbol{\mathcal{A}}_a\mathbf{z}^*(t) = \mathbf{0}_{m_a \times 1}
\end{equation}
where $\mathbf{W} \in \mathbb{R}^{m_a \times m}$ is a basis for the left null space of $\mathbf{C}_r^{\mathbf{\Pi}}$, and $\boldsymbol{\mathcal{A}}_a : m_a\times n_z$ is defined as the algebraic constraint matrix.

\subsection{Measurement error model under EIV setting} \label{subsec:2.me}
In the EIV framework, the measurements of all variables are assumed to be corrupted by additive random noise. Therefore, the measurement model relating the observed values of the variables and their true underlying values is written as:
\begin{equation}
    \label{eq:2.7}
    \mathbf{z}(t) = \mathbf{z}^*(t) + \mathbf{e_z}(t) 
\end{equation}
where $\mathbf{e_z}(t)$ is the measurement noise, that is assumed to be Gaussian white noise sequence with non-singular diagonal covariance matrix $\mathbf{\Sigma}_\mathbf{e_z}$. 

\subsection{Problem statement} \label{subsec:2.ps}
Given $N$ noisy measurements of $n_z$ variables $\{\mathbf{z}(t)\}_{t=1}^N$ from a linear DAE system \eqref{eq:2.5} under EIV setting, the identification problem investigated in this work consists of:
\begin{enumerate}
    \item Estimating the diagonal noise covariance matrix $\mathbf{\Sigma}_\mathbf{e_z}$,
    \item Estimating the number of algebraic equations $m_a$, and number of inputs $\ell$, 
    \item Identifying the algebraic constraints $\boldsymbol{\mathcal{A}}_a$ if $m_a \ne 0$,
    \item Estimating the order $n_d$ of the reduced dynamical system as well as identifying a basis of the dynamical constraint matrix $\boldsymbol{\mathcal{A}}_{d,n_d}$ in minimal realization.
\end{enumerate}

Note that the identification needs to be performed directly from $\{\mathbf{z}(t)\}_{t=1}^N$ without any further user specification such as the two prejudices: \textit{a priori} $\mathbf{y}/\mathbf{u}$ variable partition and index-$0/1$ structural pre-specification, addressed in this work.

\section{FOUNDATIONS} \label{sec:foundation}
This section reviews the subspace-based model identification using iterative PCA (SMI-IPCA)  method for identifying an EIV state-space model (index-$0$), since this forms a basis of our proposed EIV-DAE model identification approach. 

\subsection{SMI-IPCA for identifying EIV state-space models} \label{subsec:3.smi-ipca}
The identification of a linear dynamic state-space model defined in \eqref{eq:2.1} with $\mathbf{E} = \mathbf{I}_n$ in the EIV setting, can be mapped to constraint model identification of a steady-state (static) process by stacking lagged measurements of input $\mathbf{u}$ and output $\mathbf{y}$ variables. Note that SMI-IPCA framework assumes the priori knowledge of variable classification, and specifically considers $\mathbf{\Pi} = \mathbf{I}_{n_z}$. Therefore, the noise sequence in \eqref{eq:2.7} is split as $\mathbf{e_z}(t) = [\mathbf{v}(t)^\intercal,\ \mathbf{w}(t)^\intercal ]^\intercal$ associated with a noise covariance matrix $\mathbf{\Sigma}_{\mathbf{e_z}} = \mathrm{diag} \left([ \mathbf{\Sigma_v},\ \mathbf{\Sigma_w} ]\right)$. Hereafter, we denote the stacked measurements with subscript $f$, i.e., the stacking length, with the following structure of construction:
\begin{equation}
    \label{eq:3.1}
    \mathbf{y}_f(t) = \begin{bmatrix} \mathbf{y}(t)^\intercal & \mathbf{y}(t+1)^\intercal & \ldots & \mathbf{y}(t+f-1)^\intercal \end{bmatrix}^\intercal
\end{equation}
After following the similar construction for inputs $\mathbf{u}$, and noise sequences $\mathbf{w}$, and $\mathbf{v}$, \eqref{eq:2.1} can be rewritten as:
\begin{equation}
    \label{eq:3.2}
    \mathbf{y}^*_f(t) = \mathbf{\Gamma}_f\mathbf{x}(t) + \mathbf{H}_f\mathbf{u}^*_f(t)
\end{equation}
By leveraging the \textit{a priori} classification of variables into input-output sets, a lagged data matrix $\tilde{\mathbf{Z}}_f$ is constructed:
\begin{subequations}
    \label{eq:3.5}
    \begin{align}
        \tilde{\mathbf{z}}_f(t) &= \begin{bmatrix} \mathbf{y}_f(t)^\intercal & \mathbf{u}_f(t)^\intercal \end{bmatrix}^\intercal \in \mathbb{R}^{(m+\ell)f} \label{eq:3.5a} \\
        \tilde{\mathbf{Z}}_f &= \begin{bmatrix} \tilde{\mathbf{z}}_f(1) & \tilde{\mathbf{z}}_f(2) & \ldots & \tilde{\mathbf{z}}_f(N-f+1) \end{bmatrix} \label{eq:3.5b}
    \end{align}
\end{subequations}
Therefore, \eqref{eq:3.2} can be compactly written in terms of $\tilde{\mathbf{Z}}_f$ as:
\begin{subequations}
    \label{eq:3.6}
    \begin{align}
        &\left(\mathbf{\Gamma}_f^\perp\right)^\intercal \begin{bmatrix} \mathbf{I} & -\mathbf{H}_f \end{bmatrix} \tilde{\mathbf{Z}}_f^* \!=\! \mathbf{A}_d\tilde{\mathbf{Z}}_f^* \!=\! \mathbf{0}_{(mf-n)\times(N-f+1)} \label{eq:3.6a} \\
        &\mathrm{subject\ to}: \tilde{\mathbf{Z}}_f = \tilde{\mathbf{Z}}_f^* + \tilde{\mathbf{E}}_f \label{eq:3.6b}
    \end{align}
\end{subequations}
It implies that the true lagged variables in $\tilde{\mathbf{z}}_f^*(t)$ are related by $mf-n$ independent linear dynamic equations. Here, $\tilde{\mathbf{E}}_f$ is constructed from $\mathbf{e_z}(t)$. The extended observability matrix $\mathbf{\Gamma}_f$, and block Toeplitz matrix $\mathbf{H}_f$ are defined as:
\begin{align}
    \mathbf{\Gamma}_f &= \begin{bmatrix} \mathbf{C}^\intercal & (\mathbf{CA})^\intercal & \ldots & \left(\mathbf{CA}^{f-1}\right)^\intercal \end{bmatrix}^\intercal \label{eq:3.3} \\
    \mathbf{H}_f &= \mathrm{Toeplitz}\left( \begin{bmatrix} \mathbf{D} \!&\! \mathbf{CB} \!&\! \mathbf{CAB} \!&\! \ldots \!&\! \mathbf{CA}^{f-2}\mathbf{B} \end{bmatrix} \right) \label{eq:3.4} 
\end{align}


The measurement noise corrupting the lagged data vector $\tilde{\mathbf{z}}_f(t)$ is a function solely of the input and output measurement noise variances. The corresponding noise covariance matrix $\mathbf{\Sigma}_{\mathbf{e}_f}$ thus takes a block-diagonal structure:
\begin{equation}
    \label{eq:3.7}
    \mathbf{\Sigma}_{\mathbf{e}_f} = \begin{bmatrix} \mathbf{\Sigma}_{\mathbf{v}f} & \mathbf{0} \\ \mathbf{0} & \mathbf{\Sigma}_{\mathbf{w}f} \end{bmatrix}
\end{equation}
where $\mathbf{\Sigma}_{\mathbf{v}f} = \mathbf{I}_f \otimes \mathbf{\Sigma}_{\mathbf{v}}$ and $\mathbf{\Sigma}_{\mathbf{w}f} = \mathbf{I}_f \otimes \mathbf{\Sigma}_{\mathbf{w}}$, constructed from noise covariances of $\mathbf{v}(t)$ and $\mathbf{w}(t)$, respectively. Crucially, despite $\mathbf{\Sigma}_{\mathbf{e}_f}$ being of dimension $(m+\ell)f \times (m+\ell)f$, it is parameterized by only $m + \ell$ unknown noise variances.

The IPCA based method \cite{Ramnath:2023} was developed to address this identification problem, which is capable of automatically estimating system order, measurement noise variances, and system matrices. The SMI-IPCA method applies PCA on $\tilde{\mathbf{Z}}_f$ after scaling it using the lagged noise covariance matrix in \eqref{eq:3.7}, which gives the scaled lagged data matrix $\tilde{\mathbf{Z}}_{\mathbf{S}f}$ as:
\begin{equation}
    \label{eq:3.8}
    \tilde{\mathbf{Z}}_{\mathbf{S}f} = \tilde{\mathbf{Z}}_f \times \mathbf{\Sigma}_{\mathbf{e}f}^{-1/2}
\end{equation}
The property that the smallest $d$ singular values of $\tilde{\mathbf{Z}}_{\mathbf{S}f}$ should all be equal to unity as $N \to \infty$, is exploited to estimate the number of constraints $d$ using a hypothesis test to assess the equality of the smallest singular values. The right singular vectors $\mathbf{V}_d$ corresponding to $d$ unity singular values are therefore used to estimate the constraint matrix:
\begin{equation}
    \label{eq:3.9}
    \mathbf{\hat{A}}_d = \mathbf{V}_d^\intercal \times \mathbf{\Sigma}_{\mathbf{e}f}^{-1/2}
\end{equation}

Inferring from \eqref{eq:3.6a}, the system order is further estimated from the number of constraints as $\hat{n} = mf - \hat{d}$. The above model estimation step is combined with an optimization step within an iterative scheme in order to estimate the unknown measurement noise variances by minimizing the following joint likelihood of the lagged constraint residuals $\mathbf{r}_f(t)$:
\begin{equation}
    \label{eq:3.10}
    \min_{\mathbf{\Sigma}_{\mathbf{e}f}}\ (N-f+1) \log \left| \mathbf{\Sigma}_{\mathbf{r}f} \right| + \sum_{t=1}^{N-f+1} \mathbf{r}_f(t)^\intercal \mathbf{\Sigma}_{\mathbf{r}f}^{-1} \mathbf{r}_f(t) 
\end{equation}
where $\mathbf{r}_f(t) = \mathbf{\hat{A}}_d \tilde{\mathbf{z}}_f(t)$ and $\mathbf{\Sigma}_{\mathbf{r}f} = \mathbf{\hat{A}}_d \mathbf{\Sigma}_{\mathbf{e}f} \mathbf{\hat{A}}_d^\intercal$. The overall algorithm consists of two nested loops. An outer loop checks the equality of the $d$ smallest singular values, with a guessed number of constraints decremented from a maximum possible value $d_{\max} = n_zf - 1$ to a minimum possible value $d_{\min}$, that arises from the identifiability condition $d_{\min}(d_{\min} + 1) \geq 2(m + \ell)$ to ensure the unique estimation of the $m + \ell$ measurement noise variances. Nested within this is an inner loop that alternates between the model estimation step in \eqref{eq:3.8} and \eqref{eq:3.9}, and the noise variance estimation step in \eqref{eq:3.10}, iterated until convergence of the variance estimates. From $\mathbf{\hat{A}}_d$, the state-space matrices $(\mathbf{A},\mathbf{B},\mathbf{C},\mathbf{D})$ are estimated up to a rotational ambiguity through a sequence of systematic algebraic manipulations (refer to \cite{Ramnath:2023} for more details).

\section{PROPOSED METHODOLOGY} \label{sec:method}
In this section, we present a hierarchical methodology to address the linear EIV-DAE identification problem described in Section~\ref{subsec:2.ps}. We divide the entire pipeline of the proposed framework into the following three sequential steps.

\subsection{Estimation of the measurement noise variances} \label{subsec:4.noise}
The core of the proposed identification process relies on mapping the dynamic DAE formulation in \eqref{eq:2.5} into an equivalent static constraint problem, allowing the parameters to be systematically extracted via IPCA. Using $\mathbf{z}(t)$, we construct a $f_1$-lagged data matrix according to the structures defined in \eqref{eq:3.1} and \eqref{eq:3.5b}:
\begin{subequations}
    \label{eq:4.1}
    \begin{align}
        \mathbf{z}_{f_1}(t) &= \begin{bmatrix} \mathbf{z}(t)^\intercal & \mathbf{z}(t+1)^\intercal & \ldots & \mathbf{z}(t+f_1-1)^\intercal \end{bmatrix}^\intercal \label{eq:4.1a} \\
        \mathbf{Z}_{f_1} &= \begin{bmatrix} \mathbf{z}_{f_1}(1) & \mathbf{z}_{f_1}(2) & \ldots & \mathbf{z}_{f_1}(N-f_1+1) \label{eq:4.1b} \end{bmatrix}
    \end{align}
\end{subequations}
Analogous to \eqref{eq:3.2}, the subspace form of the state-space model in terms of $\mathbf{z}^*_{f_1}(t)$ can be written as follows:
\begin{equation}
    \label{eq:4.2}
    \left( \mathbf{D}_{f_1}^{\mathbf{\Pi}} - \mathbf{H}_{f_1}^{\Pi} \right) \mathbf{z}^*_{f_1}(t) = \mathbf{\Gamma}_{f_1}^{\mathbf{\Pi}} \mathbf{x}_d(t)
\end{equation}
Therefore, the identification problem can be represented as:
\begin{subequations}
    \label{eq:4.5}
    \begin{align}
        &\left( \mathbf{\Gamma}_{f_1}^{\mathbf{\Pi}^{\perp}}\right)^\intercal \left( \mathbf{D}_{f_1}^{\mathbf{\Pi}} - \mathbf{H}_{f_1}^{\Pi} \right) \mathbf{Z}^*_{f_1} = \mathbf{0}_{(mf_1-n_d)\times(N-f_1+1)} \label{eq:4.5a} \\
        &\mathrm{subject\ to}: \mathbf{Z}_{f_1} = \mathbf{Z}_{f_1}^* + \mathbf{E}_{f_1} \label{eq:4.5b}
    \end{align}
\end{subequations}
where $\left( \mathbf{\Gamma}_{f_1}^{\mathbf{\Pi}^{\perp}}\right)^\intercal \left( \mathbf{D}_{f_1}^{\mathbf{\Pi}} - \mathbf{H}_{f_1}^{\Pi} \right)$ is the dynamic constraint model, which takes minimal order form $\boldsymbol{\mathcal{A}}_{d,n_d}$ for $f_1=n_d+1$. $\mathbf{E}_{f_1}$ is constructed from $\mathbf{e_z}(t)$ in \eqref{eq:2.7} analogous to \eqref{eq:4.1a}. $\mathbf{D}_{f_1}^{\mathbf{\Pi}} = \mathbf{I}_{f_1} \otimes \mathbf{D}_r^{\mathbf{\Pi}}$, and $\mathbf{H}_{f_1}^{\Pi}$, $\mathbf{\Gamma}_{f_1}^{\mathbf{\Pi}}$ take the forms below, which are similar to the block-structure in \eqref{eq:3.3} and \eqref{eq:3.4}:
\begin{align}
    \mathbf{H}_{f_1}^{\Pi} \!&=\! \mathrm{Toeplitz}\! \left(\! \begin{bmatrix} \mathbf{0} \!&\!\! \mathbf{C}_r^{\mathbf{\Pi}}\mathbf{B}_r^{\mathbf{\Pi}} \!\!&\!\! \ldots \!&\!\! \mathbf{C}_r^{\mathbf{\Pi}}\! \left( \mathbf{A}_r^{\mathbf{\Pi}} \right)^{f_1-2}\! \mathbf{B}_r^{\mathbf{\Pi}} \end{bmatrix} \!\right) \label{eq:4.3} \\
    \mathbf{\Gamma}_{f_1}^{\mathbf{\Pi}} \!&=\! \begin{bmatrix} \left(\mathbf{C}_r^{\mathbf{\Pi}}\right)^\intercal \!\!&\!\! \left(\mathbf{C}_r^{\mathbf{\Pi}} \mathbf{A}_r^{\mathbf{\Pi}}\right)^\intercal \!\!&\!\! \ldots \!\!&\! \left(\mathbf{C}_r^{\mathbf{\Pi}} \left(\mathbf{A}_r^{\mathbf{\Pi}}\right)^{f_1-1}\! \right)^\intercal \end{bmatrix}^\intercal \label{eq:4.4}
\end{align}
Identification of a basis of the kernel $\boldsymbol{\mathcal{A}}_{d,n_d}$ along with other components as specified in Section~\ref{subsec:2.ps} starts with noise covariance estimation. As discussed in Section~\ref{subsec:3.smi-ipca}, the noise covariance matrix solely depends on the variables' measurement noise variances. Therefore, the noise covariance matrix associated with measurements of $\mathbf{Z}_{f_1}$ takes the structure $\mathbf{\Sigma}_{\mathbf{e}_{\mathbf{z},f_1}} = \mathbf{I}_{f_1} \otimes \mathbf{\Sigma}_{\mathbf{e_{z}}}$ with $n_z$ unknown noise variances. Hereafter, we essentially follow the lines of IPCA \eqref{eq:3.8}--\eqref{eq:3.10} (see Section~\ref{subsec:3.smi-ipca}) to estimate the noise variances by iteratively solving the optimization problem. $d_{\min}$ remains unchanged as the identifiability condition depends on the total number of variables present in the system.

Let the total number of constraints relating the lagged variables obtained at the end of the above iterative process be denoted by $\hat{d}_{f_1}$. It may be noted that although we obtain estimates of the noise variances and the number of constraints relating the lagged variables from this step, unlike in SMI-IPCA it is not possible to estimate the state dimension from the number of constraints, since we do not know the number of output variables.

In the second step of this framework as discussed in the following, we determine whether algebraic relations relating the variables exist among the variables.

\subsection{Discovery of algebraic constraints} \label{subsec:4.algeb}
The identification of algebraic constraints relating the augmented system variables $\mathbf{z}(t)$ is essentially a left null-space identification problem of the true unlagged data matrix $\mathbf{Z}^*$ (i.e., $f_1=0$). We apply PCA on the scaled, unlagged data matrix $\mathbf{Z}_{\mathbf{S}}$ obtained by scaling $\mathbf{Z}$ using $\mathbf{\hat{\Sigma}}_{\mathbf{e_z}}$ estimated from previous step (Section~\ref{subsec:4.noise}) as $\mathbf{Z}_{\mathbf{S}} = \mathbf{Z}\mathbf{\hat{\Sigma}}_{\mathbf{e_z}}^{-1/2}$ in order to estimate the algebraic constraint matrix $\boldsymbol{\mathcal{\hat{A}}}_a$. If $m_a$ is the dimension of the null space in which $\boldsymbol{\mathcal{\hat{A}}}_a$ lies, the smallest $m_a$ eigenvalues of the covariance matrix of $\mathbf{Z}_{\mathbf{S}}/\sqrt{N}$ should all be equal to unity. This method is known as maximum likelihood PCA (MLPCA)~\cite{Wentzell:1997}. The steps are summarized as:
\begin{subequations}
    \label{eq:4.6}
    \begin{align}
        \mathrm{svd}\left( \mathbf{Z}_{\mathbf{S}}\big/\sqrt{N} \right) &= \mathbf{U_S} \mathbf{S_S} \mathbf{V_S}^\intercal \\
        \hat{m}_a &= \mathrm{hypothesisTest}\left( \mathrm{diag}\left( \mathbf{S_S} \right)^2 \right) \\
        \boldsymbol{\mathcal{\hat{A}}}_a &= \left( \mathbf{V_S} \right)^\intercal_{\hat{m}_a} \times \mathbf{\hat{\Sigma}}_{\mathbf{e_z}}^{-1/2}
    \end{align}
\end{subequations}
where the ``$\mathrm{hypothesisTest}$'' determines $\hat{m}_a$, the number of smallest eigenvalues which are all equal to each other \cite{Ramnath:2023}. 

Furthermore, a possible set of $\hat{m}_a$ variables can be chosen as the algebraic output variables $\mathbf{z}_D(t) \in \mathbf{z}(t)$ by checking the invertibility of the submatrix $\boldsymbol{\mathcal{\hat{A}}}_{a,D} : \hat{m}_a \times \hat{m}_a$ of $\boldsymbol{\mathcal{\hat{A}}}_a$ corresponding to these variables. The remaining columns of $\boldsymbol{\mathcal{\hat{A}}}_a$ form the submatrix $\boldsymbol{\mathcal{\hat{A}}}_{a,I} : \hat{m}_a \times (n_z-\hat{m}_a)$, corresponding to the remaining variables. The implicit algebraic kernel representation yields the following regression form:
\begin{subequations}
    \label{eq:4.7}
    \begin{align}
        \boldsymbol{\mathcal{\hat{A}}}_a = \begin{bmatrix} \boldsymbol{\mathcal{\hat{A}}}_{a,D} \!&\! \boldsymbol{\mathcal{\hat{A}}}_{a,I} \end{bmatrix};& \quad \mathbf{z}^*(t) = \begin{bmatrix} \mathbf{z}^*_D(t)^\intercal \!&\! \mathbf{z}^*_I(t)^\intercal \end{bmatrix}^\intercal \\ 
        \implies \mathbf{z}^*_D(t) =& -\left( \boldsymbol{\mathcal{\hat{A}}}_{a,D} \right)^{-1} \boldsymbol{\mathcal{\hat{A}}}_{a,I} \mathbf{z}^*_I(t)
    \end{align}
\end{subequations}
where $-\big( \boldsymbol{\mathcal{\hat{A}}}_{a,D} \big)^{-1}\! \boldsymbol{\mathcal{\hat{A}}}_{a,I}$ is the algebraic regression matrix.

\subsection{Identification of differential constraints} \label{subsec:4.diff}
The differential system in \eqref{eq:2.5} intrinsically contains all the $\hat{m}_a$ algebraic relations. Therefore, the number of differential constraints along with their lagged versions associated with the $f_1$-lagged data matrix $\mathbf{Z}_{f_1}$ is computed as $\hat{d}_{f_1} - \hat{m}_af_1$. We further exploit the structural redundancy \cite{Ramanathan:2020} by constructing a second lagged data matrix $\mathbf{Z}_{f_2}$ using a stacking length of $f_2$, ($\neq f_1$). Let the number of constraints relating the lagged variables in $\mathbf{Z}_{f_2}$ be $\hat{d}_{f_2}$, which implies that the number of constraints corresponding to only the differential relations is $\hat{d}_{f_2} - \hat{m}_af_2$. The number of unique dynamic relations relating the observed variables can therefore be estimated as:
\begin{equation}
    \label{eq:4.8}
    \hat{m}_d \!=\! \frac{(\hat{d}_{f_2} \!- \hat{m}_af_2)\!-\!(\hat{d}_{f_1} \!- \hat{m}_af_1)}{f_2-f_1} \!=\! \frac{\hat{d}_{f_2} - \hat{d}_{f_1}}{f_2 - f_1} - \hat{m}_a
\end{equation}
which is obtained by equating the order of the system, that is invariant to the two different lags used. Consequently, the order of the reduced dynamic system can be estimated as $\hat{n}_d = (\hat{m}_a + \hat{m}_d)f_2 - \hat{d}_{f_2} = \hat{m}f_2-\hat{d}_{f_2}$. Finally, the number of inputs in the system is $\hat{\ell} = n_z - \hat{m}$. 

Therefore, a basis of the dynamic constraint matrix (the kernel or the null-space) $\boldsymbol{\mathcal{\hat{A}}}_{d,\hat{n}_d}$ can be obtained by applying PCA on the scaled lagged data matrix $\mathbf{Z}_{\mathbf{S}\hat{n}_d+1}$ constructed using $f=\hat{n}_d+1$, which involves the least number of lagged variables to encode all the dynamic relations in the system. If a specific input-output partition is further provided, the columns of $\boldsymbol{\mathcal{\hat{A}}}_{d,\hat{n}_d}$ can be rearranged in such a manner that $\mathbf{\Pi}=\mathbf{I}_{n_z}$, i.e., it gets associated with $\tilde{\mathbf{Z}}_{f}$ in \eqref{eq:3.5}. Given a $\mathbf{y}/\mathbf{u}$ variable partition, the sequence of algebraic manipulations applied on the rearranged constraint matrix by following Appendix 2 in \cite{Ramnath:2023} therefore leads to the estimation of the reduced system matrices $(\mathbf{\hat A}_r, \mathbf{\hat B}_r, \mathbf{\hat C}_r, \mathbf{\hat D}_r)$. This completes the presentation of the proposed algorithm.

\section{SIMULATIONS STUDIES} \label{sec:simulation}
We now demonstrate the applicability of the proposed framework through two simulation studies, that correspond to index-0 and index-1 systems, respectively. The following simulations explicitly validate the four estimation objectives outlined in Section \ref{subsec:2.ps}.

\subsection{Index-0 ODE system: Fourth-order 2$\times$2 dynamic process} \label{susec:5.wang}
We consider the $2\times2$ fourth-order dynamic process drawn from \cite{Wang:2002}, as described in \eqref{eq:5.1}. As input $\mathbf{u}^*$, two full-band RBS signals of length $N=6095$ are used to generate the true values of $\mathbf{y}^*$. We construct $\mathbf{z}^*$ as $[\mathbf{u}^{*\intercal},\ \mathbf{y}^{*\intercal}]^\intercal$, which are corrupted using Gaussian noises of variances $\mathrm{diag}(\mathbf{\Sigma}_{\mathbf{e_z}}) = [0.100,\ 0.099,\ 6.826,\ 15.577]^\intercal$ that correspond to a signal-to-noise ratio (SNR) of $10$.
\begin{subequations} \label{eq:5.1}
    \begin{align}
        \mathbf{x}(t+1) =\ &\text{diag}\left(\begin{bsmallmatrix} 0.67 & 0.67 \\ -0.67 & 0.67 \end{bsmallmatrix}, \begin{bsmallmatrix} -0.67 & -0.67 \\ 0.67 & -0.67 \end{bsmallmatrix}\right) \mathbf{x}(t)\ + \nonumber \\ &\begin{bsmallmatrix} 0.6598 & 1.9698 & 4.3171 & -2.6436 \\ -0.5256 & 0.4845 & -0.4879 & -0.3416 \end{bsmallmatrix}^\intercal \mathbf{u}^*(t) \\
        \mathbf{y}^*(t) =\ &\begin{bsmallmatrix} -0.5749 & 1.0751 & -0.5225 & 0.1830 \\ 2.4027 & 0.7543 & -0.2159 & 0.0982 \end{bsmallmatrix} \mathbf{x}(t)\ + \nonumber \\ &\begin{bsmallmatrix} -0.7139 & -0.1174 \\ 0.3131 & -0.2876 \end{bsmallmatrix} \mathbf{u}^*(t)
    \end{align}
\end{subequations}

\begin{figure}[tb]
\centerline{
    \includegraphics[width=0.99\columnwidth]{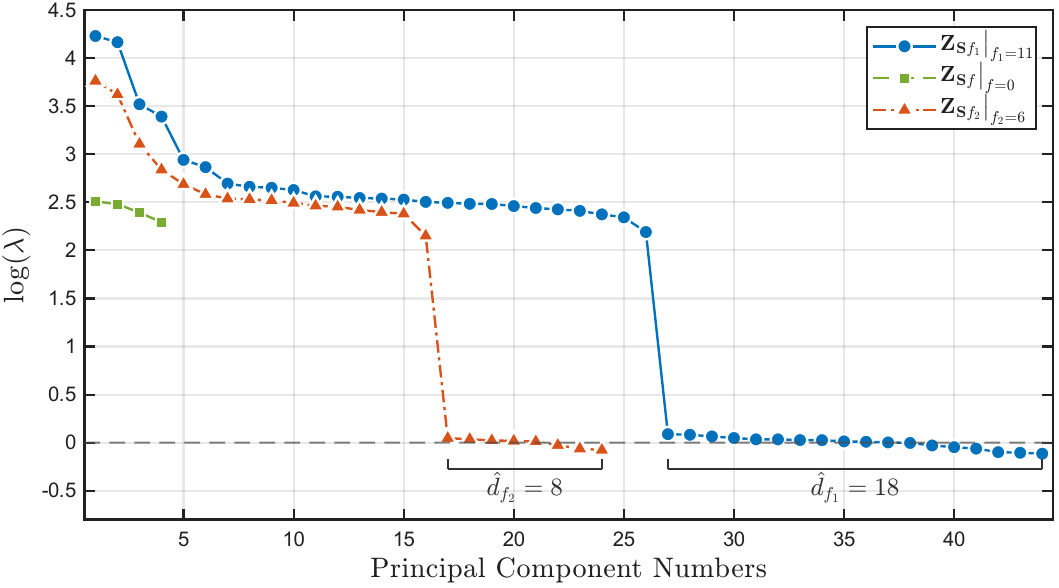}
}
\caption{Scree plots associated with the index-0 system. Eigenvalues, that are close to unity, are along the $\log(1)$ line. $\lambda$ denotes the eigenvalues of the sample covariance matrix associated with the scaled lagged/unlagged data matrices used as the legends at the top-right corner.}
\label{fig_1}
\end{figure}

As the first step of the proposed algorithm, in order to estimate the noise variances, we choose a stacking length $f_1=11$. Starting with $d_{\max} = 43$, the hypothesis test fails to reject the null hypothesis when $\hat{d}_{f_1}=18$ as also reported in Fig.~\ref{fig_1}. The point estimates of the noise variances are $\mathrm{diag}(\mathbf{\hat{\Sigma}}_{\mathbf{e_z}}) = [0.099,\ 0.091,\ 6.909,\ 15.656]^\intercal$, which are clearly very close to the respective true values. The unlagged data matrix $\mathbf{Z}$ is scaled using $\mathbf{\hat{\Sigma}}_{\mathbf{e_z}}$ as discussed in the second step of the algorithm (Section~\ref{subsec:4.algeb}). The hypothesis test yields $0$ unity eigenvalues, which is evident from Fig.~\ref{fig_1}, implying that there is no algebraic constraint relating the variables in $\mathbf{z}$. Therefore, it correctly identifies the index ($=0$) of the system. 

In the third step, a second lagged data matrix is constructed with a lag value $f_2=6$, and the number of linear constraints relating the lagged variables in $\mathbf{Z}_{f_2}$ comes out to be $\hat{d}_{f_2}=8$ (see Fig.~\ref{fig_1}). Since there is no algebraic variables in the system, the number of differential output variables are $\hat{m}_d=(8-18)/(6-11) = 2$, and the order of the dynamic subsystem becomes $\hat{n}_d=2\times 6-8 = 4$, both of which are correctly estimated by the algorithm. Therefore, there are $\hat{\ell} = 2$ input variables. We further construct a lagged data matrix using $f=\hat{n}_d+1 = 5$, and obtain an estimate of the dynamic constraint matrix $\boldsymbol{\mathcal{\hat{A}}}_{d,5} \in \mathbb{R}^{6 \times 20}$ by performing PCA. We adopt the original partition of variables used to generate the data to rearrange the columns of $\boldsymbol{\mathcal{\hat{A}}}_{d,5}$ such that $\mathbf{\Pi}=\mathbf{I}_4$ as discussed in Section~\ref{subsec:4.diff}, and estimate the state-space matrices up to a rotational ambiguity. We therefore compare the poles-zeros of the estimated and true state-space models. The results are reported in Table~\ref{tab_1}, that are averaged over $200$ runs of bootstrapping, and the $95\%$ confidence intervals (CIs) imply the estimates to be unbiased.

\begin{table}[tb]
\caption{Estimated poles and zeros of the index-0 system}
\begin{center}
    \renewcommand{\arraystretch}{1.05}
    \begin{tabular}{|c|c|c|c|}
    \hline
    \textbf{True poles} ($p$) & $\mu(\hat{p})$ & $\sigma\left( \Re(\hat{p}) \right)$ & $\sigma\left( \Im(\hat{p}) \right)$ \\
    \hline
    $-0.670 \pm 0.670i$ & $-0.671 \pm 0.669i$ & $0.0007$ & $0.0007$ \\
    $0.670 \pm 0.670i$ & $0.670 \pm0.668i $ & $0.0011$ & $0.0009$ \\
    \hline
    \hline
    \textbf{True zeros} ($z$) & $\mu(\hat{z})$ & $\sigma\left( \Re(\hat{z}) \right)$ & $\sigma\left( \Im(\hat{z}) \right)$ \\
    \hline
    $-2.602 \pm 1.412i$ & $-2.780 \pm 1.287i$ & $0.0987$ & $0.1019$ \\
    $-1.085$ & $-1.087$ & $0.0174$ & $0$ \\
    $3.067$ & $2.848$ & $0.0817$ & $0$ \\
    \hline
    \end{tabular}
\label{tab_1}
\end{center}
\end{table}

\subsection{Index-1 DAE system: Two-tank liquid-level process} \label{subsec:5.tank}
\begin{figure}[tb]
\centerline{
    \includegraphics[width=0.65\columnwidth]{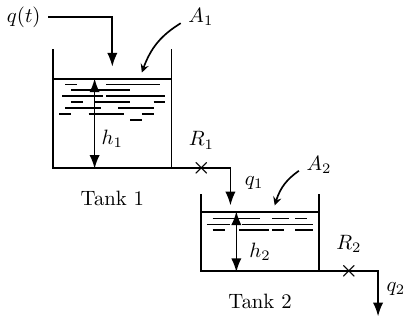}
}
\caption{Schematic of a non-interacting two-tank liquid level system.}
\label{fig_2}
\end{figure}

The non-interacting liquid level system \cite{LeBlanc:2009} consisting of two tanks, is schematically shown in Fig.~\ref{fig_2}. Outlet from tank 1 discharges directly into tank 2. Assuming incompressible flow and uniform tank geometry, the system exhibits linear resistance characteristics. Crucially, the flow through valves $R_1, R_2$ is determined only by the respective upstream levels, meaning $h_2$ exerts no back-pressure on the preceding stage. The resulting first-principles model, based on the conservation of mass, yields:
\begin{equation}
    \label{eq:5.2}
    q(t) - q_1(t) = A_1\frac{dh_1(t)}{dt}; \quad q_1(t) - q_2(t) = A_2\frac{dh_2(t)}{dt}
\end{equation}
whereas the flow-head relationships for the two linear resistances are defined as:
\begin{equation}
    \label{eq:5.3}
    q_1(t) = \frac{h_1(t)}{R_1}; \quad q_2(t) = \frac{h_2(t)}{R_2}
\end{equation}

The system is excited with a full-band RBS input $q(t)$ around the nominal operating condition with cross-sectional areas $A_1=1.1\ \text{m}^2$, $A_2=2.4\ \text{m}^2$, and resistances of the outgoing liquid $R_1=0.75\ \text{s/m}^2$, and $R_2=1.5\ \text{s/m}^2$, in order to obtain $N=6095$ true values of $\mathbf{z} = [q,\ q_1,\ q_2,\ h_1,\ h_2]^\intercal$ with a sampling interval of $1$s. Gaussian white noise of variances $\mathrm{diag}(\mathbf{\Sigma}_{\mathbf{e_z}}) = [0.100,\ 0.053,\ 0.011,\ 0.030,\ 0.024]^\intercal$ corresponding to SNR $=10$ are added to the noise-free data.

\begin{figure}[tb]
\centerline{
    \includegraphics[width=0.99\columnwidth]{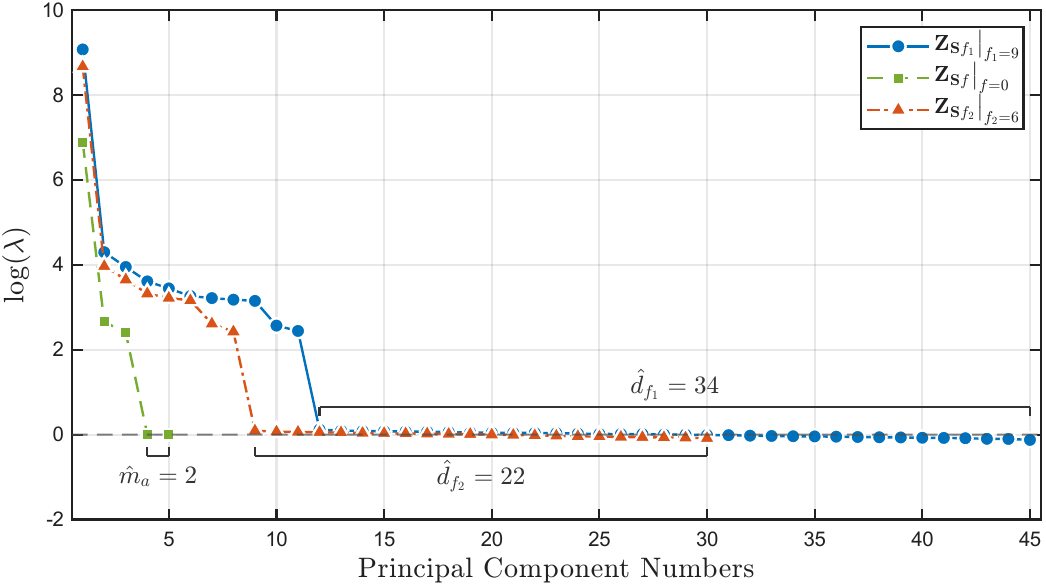}
}
\caption{Scree plot associated with the index-1 system.}
\label{fig_3}
\end{figure}

We choose a stacking length of $f_1=9$ in the first step, and the total number of constraints $\hat{d}_{f_1}$ relating the variables in the $f_1$-lagged data matrix $\mathbf{Z}_{f_1}$ comes out to be $34$ as also reported in Fig.~\ref{fig_3}. Simultaneously, we obtain accurate point estimates of the noise variances $\mathrm{diag}(\mathbf{\hat{\Sigma}}_{\mathbf{e}_{\tilde{z}}}) = [0.099,\ 0.056,\ 0.010,\ 0.031,\ 0.024]^\intercal$. Subsequently, we scale the unlagged data matrix, and hypothesis test indicates that the smallest $2$ eigenvalues are unity (see Fig.~\ref{fig_3}), implying the existence of $\hat{m}_a=2$ algebraic constraints. Among the ${}^5C_{\hat{m}_a}$ possible choices, we observe $h_1,h_2$ variables to be the most robust choice as algebraic output variables with a condition number of $1.149$ associated with the dependent submatrix. Consequently, MLPCA yields the following $2$ algebraic equations in regression form, where the coefficients, averaged over $200$ bootstrap runs, are reported along with their $95\%$ CIs.
\begin{subequations}
    \label{eq:5.4}
    \begin{align}
        h_1^*(t) &= \underset{\pm 0.007}{-0.005}q^*(t) + \underset{\pm 0.009}{0.758}q_1^*(t) - \underset{\pm 0.006}{0.003}q_2^*(t) \\
        h_2^*(t) &= \underset{\pm 0.003}{0.0001}q^*(t) - \underset{\pm 0.008}{0.007}q_1^*(t) + \underset{\pm 0.006}{1.505}q_2^*(t)
    \end{align}
\end{subequations}

We further choose $f_2=6$ to construct a second lagged data matrix $\mathbf{Z}_{f_2}$. Eigenvalue analysis on $\mathbf{Z}_{f_2}$ results in $\hat{d}_{f_2}=22$ linear constraints encoding both the differential and algebraic constraints. Following \eqref{eq:4.8}, the number of dynamical output variables is estimated to be $\hat{m}_d = (22-34)/(6-9) - 2 = 2$, whereas the reduced system order is estimated as $\hat{n}_d = 4\times 6 - 22 = 2$. Finally in order to obtain the global constraint model, we use $f=\hat{n}_d+1=3$ to construct the data matrix, from which we obtain $\boldsymbol{\mathcal{\hat{A}}}_{d,3} \in \mathbb{R}^{10\times 15}$. Since, we cannot compare the estimated constraint matrix with the true model, we use the partition of input and output variables specific to the data generation process and rearrange the columns of $\boldsymbol{\mathcal{\hat{A}}}_{d,3}$ such that the first $12$ columns correspond to output variables and their lagged versions, whereas the last $3$ columns associate with the lagged inputs. Following the algebraic manipulations in \cite{Ramnath:2023}, we obtain the estimate of the state-space matrices up to a rotational ambiguity, from which the system poles come out to be $[0.764\pm0.008,\ 0.286\pm0.012]$ which are very close the true system poles $[0.758,\ 0.298]$, and zeros of this system is an empty set. Therefore, we conclude that the algorithm accurately estimates all the system parameters.

\section{CONCLUSION} \label{sec:conclusion}
This paper has proposed an EIV subspace identification framework for linear discrete-time DAE systems that eliminates two fundamental prejudices: the \textit{a priori} input-output variable partition and the structural index pre-specification. Simulation results confirm that the algorithm correctly estimates all system parameters for both index-0 and index-1 DAE systems. The proposed identification strategy can be used for causal discovery in linear descriptor systems.  Future research directions include extending the framework to systems containing process noise and higher-index DAEs.


\end{document}